\documentclass[prd,aps,twocolumn,nofootinbib,preprintnumbers,superscriptaddress,preprintnumbers,balancelastpage,longbibliography]{revtex4-2}
\pdfoutput=1
\usepackage{amsmath,mathtools,physics,xfrac}
\usepackage{graphicx}
\usepackage{afterpage}
\usepackage{float}
\usepackage{subfigure}
\usepackage{rotating}
\usepackage{multirow}
\usepackage{tabularx}
\usepackage{booktabs}
\usepackage{fancyhdr}
\usepackage[utf8]{inputenc}
\usepackage{theorem}
\usepackage{moreverb}
\usepackage{euscript}
\usepackage{psfrag}
\usepackage{slashed}
\usepackage{mathtools}
\usepackage{makecell}
\usepackage{adjustbox}
\usepackage{dcolumn}
\usepackage{bm}
\usepackage[dvipsnames]{xcolor}
\usepackage{graphics}
\usepackage{hyperref}
\usepackage{lettrine}

\input Zallman.fd

\LettrineTextFont{\itshape}
\hypersetup{
     colorlinks   = true,
     citecolor    = cyan!75!Blue,
     urlcolor     = cyan!75!Blue,
     linkcolor    = cyan!75!Blue
}

\newcommand{\be}{\begin{equation}}
\newcommand{\ee}{\end{equation}}
\newcommand{\bea}{\begin{eqnarray}}
\newcommand{\eea}{\end{eqnarray}}

\definecolor{nice}{rgb}{0.8,0, 0.8}

\begin{document}

\preprint{SLAC-PUB-17752}

\title{Dark Matter Scattering Constraints from Observations of Stars Surrounding Sgr A*}
\author{Isabelle John}
\thanks{\href{mailto:isabelle.john@fysik.su.se}{isabelle.john@fysik.su.se}, \href{http://orcid.org/00000-0003-2550-7038}{ORCID: 0000-0003-2550-7038}}
\affiliation{Stockholm University and The Oskar Klein Centre for Cosmoparticle Physics,  Alba Nova, 10691 Stockholm, Sweden}
\author{Rebecca K. Leane}
\thanks{\href{mailto:rleane@slac.stanford.edu}{rleane@slac.stanford.edu}, \href{http://orcid.org/0000-0002-1287-8780}{ORCID: 0000-0002-1287-8780}}
\affiliation{Particle Theory Group, SLAC National Accelerator Laboratory, Stanford, CA 94035, USA}
\affiliation{Kavli Institute for Particle Astrophysics and Cosmology, Stanford University, Stanford, CA 94035, USA}\author{Tim Linden}
\thanks{\href{mailto:linden@fysik.su.se}{linden@fysik.su.se}, \href{http://orcid.org/0000-0001-9888-0971}{ORCID: 0000-0001-9888-0971}}
\affiliation{Stockholm University and The Oskar Klein Centre for Cosmoparticle Physics,  Alba Nova, 10691 Stockholm, Sweden}

\begin{abstract}
High resolution infrared data has revealed several young stars in close proximity to Sgr A*. These stars may encounter extremely high dark matter densities. We examine scenarios where dark matter scatters on stellar gas, accumulates in stellar cores, and then annihilates. We study the stars S2, S62, S4711 and S4714 and find three observable effects. First, dark matter interactions can inhibit \emph{in situ} star-formation close to Sgr A*, favoring scenarios where these stars migrate into the Galactic Center. Second, dark matter interactions can delay main sequence evolution, making stars older than they appear. Third, very high dark matter densities can inject enough energy to disrupt main sequence stars, allowing S-star observations to constrain the dark matter density near Sgr A*.
\end{abstract}

\maketitle

\lettrine{S}{tars} are among the densest objects in the Universe, making them unique targets for dark matter searches~\cite{1989ApJ...338...24S,Fairbairn:2007bn, Scott:2008ns, Iocco:2008xb, Hooper:2010es, Iocco:2012wk, PhysRevLett.55.257, Casanellas:2014zia, Lopes_2019, Hassani:2020uhd, Leane:2022hkk, Raen:2020qvn, Lopes:2020dau,Croon:2023trk}. Within this context, their luminosities are both a blessing and a curse; stars are observable at large distances and in diverse dark matter environments. However the vast majority of their flux stems from baryonic effects that are unrelated to any feeble dark matter interactions. Thus, stellar probes of dark matter either focus on very local sources, for which precise observations are possible (e.g., the Sun~\cite{Batell:2009zp, Schuster:2009fc, Bell:2011sn, PhysRevD.93.115036, Allahverdi:2016fvl, Leane:2017vag, Arina:2017sng, HAWC:2018szf, Nisa:2019mpb, Niblaeus:2019gjk, Cuoco:2019mlb, Mazziotta:2020foa, Bell:2021pyy, Chauhan:2023zuf}), or on stars in special targets where dark matter interactions may be enhanced (e.g., dwarf galaxies, the Galactic Center, or in the early universe)~\cite{1989ApJ...338...24S,Fairbairn:2007bn, Iocco:2008xb,Freese:2008hb, Scott:2008ns, Sivertsson:2010zm, 2010ApJ...716.1397F, Freese:2015mta, Hassani:2020zvz, Ilie:2020nzp, Leane:2021ihh, Ellis:2021ztw, 2021arXiv210603398H, Acevedo:2023xnu, Qin:2023kkk}. 

Over the last decade, advances in ultra-high resolution infrared observations have unlocked our ability to detect stars that lie extremely close to Sagittarius A* (Sgr A*), the supermassive black hole at the center of our Galaxy~\cite{Ghez:2003qj, Martins:2007rv,  
 Ghez:2008ms, Gillessen:2008qv, 2010RvMP...82.3121G, 
2017ApJ...847..120H, Pei_ker_2020}. These stars, known as the S-stars, have relatively high masses (likely a selection effect due to the optical attenuation in the Galactic Center), and high eccentricities of $e \approx 0.9$. The origin of the S-stars is debated, and may be due to {\it in situ} formation, gas infall, or binary/triple interactions after star formation~\cite{Mapelli:2008dz, 2014ApJ...787L..14W, Mapelli:2015uia}.

\begin{figure}
\centering
\includegraphics[width=0.495\textwidth]{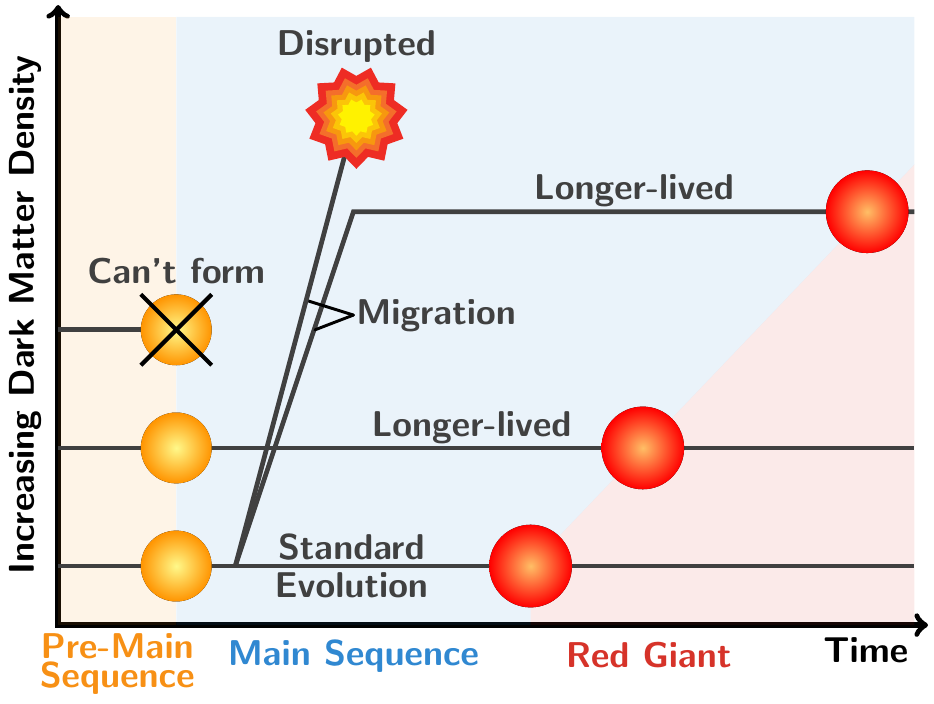}
\caption{Schematic of our findings on the impact of dark matter in stars. For low dark matter densities, stellar evolution follows the standard scenario. As the dark matter density increases, stars that form \textit{in situ} are partially powered by dark matter annihilation and become longer-lived, and for large enough dark matter densities instead fail to form. Main sequence stars formed in a lower density region can migrate to a high density region and, as the dark matter density increases, become longer-lived, or disrupted when dark matter annihilation overpowers nuclear fusion.}
\label{fig:schematic}
\end{figure}

The extreme proximity of these stars to Sgr A* (${ \lesssim 1000 }$~AU), in particular at pericenter, provides a unique opportunity to study dark matter in a region where the dark matter density may be extremely large. In such a scenario, the total energy injection from capture and subsequent annihilation can compete with nuclear fusion and significantly affect the energetics of the star itself~\cite{1989ApJ...338...24S,Fairbairn:2007bn, Scott:2008ns}. Notably, systems near the Galactic Center might even become ``WIMP burners"~\cite{1989ApJ...338...24S,Fairbairn:2007bn, Scott:2008ns}, where the stellar luminosity is driven primarily by dark matter annihilation, making these systems a near-universe version of ``dark stars"~\cite{Iocco:2008xb,Freese:2008hb, Sivertsson:2010zm, Freese:2015mta}.

In this \emph{paper}, we use S-star observations to constrain annihilating dark matter being captured by stars. Our results depend sensitively on the dark matter density near Sgr~A*, as shown in Fig.~\ref{fig:schematic}, and can be divided into three regimes. First, for relatively standard dark matter density profiles, dark matter annihilation can prevent \textit{in situ} star formation near Sgr~A*~\cite{Scott:2008ns}, providing support for astrophysical models where these stars migrate from larger radii. Second, dark matter annihilation can delay main sequence stellar evolution, allowing these stars to survive up to twice as long as expected for their mass. Third, steeply peaked dark matter profiles can inject so much energy that main sequence stars are entirely disrupted, allowing S-star observations to strongly constrain the dark matter scattering cross section for well-motivated dark matter models.

Our paper is organized as follows. In Sec.~\ref{sec:meth}, we qualitatively summarize the setup of our investigation. In Sec.~\ref{sec:darkints}, we describe the input from dark matter interactions, including dark matter profiles and capture rates. We then detail the measured properties of our S-cluster stars in Sec.~\ref{sec:obs}, and describe the stellar modeling and evolution process we simulate with \texttt{MESA}. We discuss our results in Sec.~\ref{sec:results} and conclude in Sec.~\ref{sec:conc}.

\section{Physical Picture}
\label{sec:meth}

\subsection{Dark Matter-Stellar Interactions}

Our study focuses on classes of WIMP-like (Weakly Interacting Massive Particle) dark matter models that have both a nuclear scattering cross section and an annihilation cross section. We generally consider model classes that are motivated by thermal-relic considerations~\cite{Steigman:2012nb,Leane:2018kjk}, but note that the high concentration of dark matter within stellar cores implies that the exact strength of the annihilation cross-section plays a negligible role in determining whether dark matter that has been accreted within the star subsequently annihilates.

The sequential steps of our dark matter/stellar interactions are: 

\begin{enumerate}
    \item Dark matter surrounding Sgr A* encounters a given star. The dark matter particle scatters with a baryon, robbing energy from the dark matter and trapping it within the star.
    \item Over a short time, further interactions can occur and decrease the dark matter kinetic energy, causing it to primarily fall into the stellar core~\cite{1987ApJ...321..571G,1992ApJ...387...21G,Bramante:2017xlb,Leane:2022hkk,Leane:2023woh}.
    \item The rising dark matter density in the stellar core induces efficient dark matter annihilation even for small annihilation cross-sections. Because the annihilation rate goes as the dark matter density squared, and removes two dark matter particles, while nuclear scattering only adds a single particle, the annihilation and capture rates generally can easily and rapidly reach an equilibrium.
    \item The dark matter annihilation energy at the stellar core, primarily in the form of charged particles and $\gamma$-rays, cannot escape (assuming short-range interactions). This energy instead thermalizes and produces a significant new heat source for the star, which competes with stellar fusion and affects the subsequent stellar evolution.
\end{enumerate}

\subsection{Overview of Stellar Evolution}

In order to understand the impact of dark matter heating on stellar evolution, we give a brief overview of the standard stellar evolution picture. A useful visualization of the different evolutionary stages are Hertzprung-Russell (HR) diagrams, that show the stellar luminosity against the effective stellar temperature (on a reversed axis). A newly forming star (pre-main sequence star) starts out at low temperatures and low luminosities (i.e. at the bottom right of the HR diagram; the exact position depends on the initial stellar mass). As this star gravitationally contracts further, its luminosity decreases, which results in an approximately vertical downwards motion on the HR diagram, called the Hayashi track. Once the forming star has contracted enough to ignite hydrogen fusion in its core, temperature and luminosity increase rapidly, (i.e. the star moves towards the left side of the HR diagram), until the star reaches its main sequence stage.

Once on the main sequence, the star undergoes hydrogen fusion in the core in order maintain hydrostatic equilibrium, and gravitational forces and radiation pressure balance each other out so that the star is stable and its luminosity and temperature are steady. The star spends most of its lifetime on the main sequence, but eventually the hydrogen in the core is exhausted. Instead, the star undergoes hydrogen fusion in a shell around the core. This causes the star to expand and results in a temperature decrease and luminosity increase (i.e. the star moves towards the upper right of the HR diagram, away from the main sequence). When the hydrogen in the shell is depleted as well, Helium burning ignites in the stellar core (indicated by a so-called Helium flash), and the stellar temperature increases once again. The star expands to hundreds of times its original main-sequence size, and becomes a red giant. The subsequent evolutionary stages depend on the star's mass; lighter stars cool and become white dwarves, while more massive stars turn into neutron stars. However, in this study, we only focus on stellar evolution before and on the main sequence.

\section{Dark Matter Interactions}
\label{sec:darkints}

\subsection{Dark Matter Profiles}

First, to calculate the rate in which dark matter is captured along the stellar orbit, and hence the dark matter annihilation rate, we need to know how dark matter is distributed in our Galaxy. We utilize a dark matter density profile that is based on a standard Navarro, Frenk, White (NFW) profile~\cite{Navarro:1995iw} with a free profile index $\gamma$, since the exact profile is not known, 
\begin{equation}\label{eq: DM profile}
\rho(r) = \rho_\text{scale}\left(\frac{r}{R_\text{scale}}\right)^{-\gamma} \left(1 + \frac{r}{R_\text{scale}}\right)^{\gamma-3},
\end{equation}
with the scale density
\begin{equation}\label{eq: scale density}
\rho_\text{scale} = \rho_\odot \left(\frac{R_\odot}{R_\text{scale}}\right)^\gamma\left(1 + \frac{R_\odot}{R_\text{scale}}\right)^{3-\gamma},
\end{equation}
where $r$ is the distance from the Galactic Center, ${ R_\text{scale} = 20}$~kpc is taken as the scale radius, ${ \rho_\odot = 0.4 }$~GeV/cm$^3$~\cite{Read:2014qva} is the local dark matter density, ${ R_\odot=8.5}$~kpc is the approximate distance of the Sun from the Galactic Center~\cite{Read:2014qva, 2021A&A...647A..59G}, and $\gamma$ is the profile index, where $\gamma = 1$ returns the standard NFW profile~\cite{Navarro:1995iw}. To bracket a wide range of possible dark matter profiles, we consider values for $\gamma$ that span from 0.5 --- 1.5~\cite{Gnedin:2003rj, Portail_2015, Iocco:2016itg, Hooper:2016ggc, Zuriaga-Puig:2023imf}.

To test a range of dark matter profiles, we will also consider a separate case where the dark matter profile has a density spike in the Galactic Center. Such profiles have been proposed as dark matter would be accreted by the black hole, strongly increasing the dark matter density towards the inner Galaxy~\cite{Gondolo:1999ef}. Here, we follow the spike model from~\cite{Lacroix:2018zmg}, 
\begin{equation}\label{eq: DM profile with spike}
\rho_\text{spike}(r) = \begin{cases}
0 & r < 2R_S \\
\rho(R_\text{spike})\left(\frac{r}{R_\text{spike}}\right)^{-\gamma_\text{spike}} & 2R_S \le r < R_\text{spike} \\
\rho(r) & r \ge R_\text{spike}
\end{cases}
\end{equation}
where $R_\text{spike}$ and $\gamma_\text{spike}$ are the spike parameters. In our default model we take $\gamma_\text{spike} = 7/3$ and ${ R_\text{spike} = 10} $~pc, but will also consider values for $R_\text{spike}$ between 0.01 and 100~pc~\cite{McMillan_2016, Lacroix:2018zmg, Zuriaga-Puig:2023imf}. The dark matter profile of the halo $\rho(r)$ is given by Eq.~(\ref{eq: DM profile}), and $R_S$ is the Schwarzschild radius of the black hole given by
\begin{equation}\label{eq: Schwarzschild radius}
R_S = \frac{2\,GM_\text{BH}}{c^2},
\end{equation}
where $M_\text{BH} \sim 4\times 10^6 M_\odot$ is the mass of Sgr A*, $G$ is the gravitational constant, and $c$ is the speed of light.

The spiked dark matter density can in fact become so high that the dark matter particles annihilate away, resulting in a flattening of the dark matter profile towards the Galactic Center. This saturation density is
\begin{equation}\label{eq: DM density saturation}
\rho_\text{saturation} = \frac{m_\chi}{\langle\sigma v\rangle\, t_\text{BH}},
\end{equation}
where $t_\text{BH}$ is the age of Sgr A*, $t_\text{BH} \sim 10^{10}$~yr. For example, for a dark matter particle with mass ${ m_\chi = 1 }$~TeV and the velocity averaged dark matter annihilation cross section ${ \langle\sigma v\rangle \sim 10^{-26} }$~cm$^3$/s, the maximum density is $\rho_\text{saturation} \sim 3.2\times10^{11}$~GeV/cm$^3$.

\subsection{Dark Matter Capture Rate in the Star}
Given the dark matter density distributions above, we can calculate the dark matter capture rate for a star along its orbit around Sgr A*. The dark matter particles scatter inside the star, lose energy, and become trapped. Here we describe the calculation of the dark matter capture rate and the conversion of capture rates to interaction cross sections between the dark matter particles and the nucleons in the star~\cite{Bramante:2017xlb, Ilie:2020vec}, following Ref.~\cite{Leane:2023woh}.

The capture rate after $N$ or fewer scatters is given by~\cite{Leane:2023woh}
\begin{equation}
    C_N = \pi R^2  p_N(\tau) \, \sum_{i=1}^{ N} \int_0^\infty f(u)  u  \left( 1 + w^{-2} \right) \, g_i(u)\, du  \,,
    \label{eq: cn capture rate}
\end{equation}
where $f(u)$ is the dark matter (DM) velocity distribution, and at exactly the $N$th scatter,
\begin{equation}
    g_N(u) = 1 - \frac{1}{\beta}  + \frac{1}{ \beta^N \left( 1+ w^2 \right) } \log{ \left[\frac{1}{1 - \beta}\right]}^{N-1}\,,
    \label{eq:gnexact}
\end{equation}
with $w=u/v_{\rm esc}$ where $u$ is the velocity of the incoming dark matter particle, $v_{\rm esc}$ is the stellar escape velocity, and
\begin{equation}
    \beta=\frac{4 m_{\rm SM}\, m_{\chi}}{(m_{\rm SM} + m_{\chi})^2},
\end{equation}
where $m_\chi$ is the dark matter mass, and $m_{\rm SM}$ is the mass of the standard model target particles in the star.
Finally, in Eq.~(\ref{eq: cn capture rate}), the probability of a dark matter particle undergoing $N$ scatters in the star is given by
\begin{equation}\label{eq: scattering probability}
p_N(\tau) = 2 \int_0^1 \frac{ye^{-y\tau}\left(y\tau\right)^N}{N!} dy,
\end{equation}
where $y$ takes into account the incident angle of the dark matter particle. The dark-matter--nucleon scattering cross section $\sigma_{\chi N}$ enters the formalism through the optical depth,
\begin{equation}\label{eq: optical depth}
\tau = \frac{3}{2} \frac{\sigma_{\chi N}}{\sigma_\text{tr}},
\end{equation}
where the transition cross section that marks the switch between the single and multi-scatter capture regime is
\begin{equation}\label{eq: transition cross section}
\sigma_\text{tr} = \frac{\pi R^2}{N_A},
\end{equation}
with $R$ the stellar radius and $N_A$ the number of stellar atoms of mass $A$. We consider main sequence stars and thus assume as a simplifying case that the stars are entirely composed of hydrogen.

The total capture rate is then found by summing Eq.~(\ref{eq: cn capture rate}) up to a maximal number of scatters $N_\text{max}$, as 
\begin{equation}
    C_\text{total} = \sum_{N=1}^{N_\text{max}} C_N .
\end{equation}
We implement the capture formalism using the \texttt{Asteria} package~\cite{Leane:2023woh}, which also includes reflection of light dark matter in the strong interaction regime (which is based on simulations in Ref.~\cite{Leane:2023woh} which we have not detailed here). As per Ref.~\cite{Leane:2023woh}, we do not include thermal effects of the stellar nuclei, however this choice is conservative as their inclusion overall tends to increase the capture rate, see e.g.~Ref.~\cite{Garani:2017jcj}.

We will describe our stars of interest shortly, but note that the orbits of our stars are highly eccentric, which means that the dark matter capture rate varies strongly between the pericenter and apocenter distances of the star. At pericenter, the stellar velocities reach several percent of the speed of light. In our computations, we use the average dark matter capture rate and density integrated along the orbit. For the kinematics of the interaction, the dark matter velocity will be negligible compared to the stellar velocity which is very fast. We therefore use the stellar velocity only in the kinematics.

\subsection{Dark Matter Equilibrium and Annihilation}
Inside the star, the dark matter annihilates and thus provides extra power to the star additionally to nuclear fusion. The dark matter capture and annihilation process can fall into equilibrium, where the two rates are equal. The equilibrium time scale can be estimated from
\begin{equation}\label{eq: equilibrium time scale}
t_\text{eq} = \sqrt{\frac{4}{3}\frac{\pi R^3}{C_\text{total} \langle\sigma v\rangle}},
\end{equation}
which gives $t_\text{eq} \sim \mathcal{O}(100 - 1000)$~yr for the dark matter mass range and annihilation cross sections relevant here. This is longer than the orbital period of our stars of interest ($t_\text{period} \sim \mathcal{O}(10)$~yr, see Tab.~\ref{tab: star parameters}),  and so we assume that the dark matter annihilation rate is equal to the dark matter capture rate throughout our analysis. Furthermore, while the capture rate changes along the orbit according to the stellar velocity and dark matter density, we do not expect any stellar observable to change significantly over the course of the stellar orbit, given that both the relaxation scale for dark matter capture, and the stellar photon diffusion timescale typically exceed the $\sim$~10~yr orbital cycle. 

\section{S-Star Observations and Modeling}
\label{sec:obs}

\begin{table*}[t!]
\centering
\renewcommand{\arraystretch}{1.5}
\begin{tabular}{|c|c|c|c|c|}
\hline \rule{0pt}{3ex}

                       & \textbf{S2}      & \textbf{S62}      & \textbf{S4711}     & \textbf{S4714}    \\
\hline \rule{0pt}{3ex}
 Age [Myr]             & 6.6$^{+3.4}_{-4.7}$ & --  & -- & -- \\ \rule{0pt}{3ex}
 $M$ [$M_\odot$]       & 13.6$^{+2.2}_{-1.8}$ & 6.1 & 2.2 & 2.0 \\ \rule{0pt}{3ex}
 $R$ [$R_\odot$]       & 5.53$^{+1.77}_{-0.79}$  & --  & -- & -- \\ \rule{0pt}{3ex}
 $T_\text{eff}$ [K]    & 28\,513$^{+2388}_{-2923}$ & --  & -- & -- \\ \rule{0pt}{3ex}
 $\log{L/L_\odot}$     & 4.35$^{+0.18}_{-0.19}$  & -- & -- & -- \\ \rule{0pt}{3ex}
 $r_p$ [AU]            & 119.3 $\pm$ 0.3 & 17.8 $\pm$ 7.4 & 143.7 $\pm$ 18.8 & 12.6 $\pm$ 9.3 \\ \rule{0pt}{3ex}
 $r_a$ [AU]            & 1949.9 $\pm$ 2.8  & 1462.4 $\pm$ 11.0  & 1094.7 $\pm$ 28.7  & 1670 $\pm$ 10 \\ \rule{0pt}{3ex}
 $v_p$ [km/s], \% c    & 7582 $\pm$ 8, 2.527 $\pm$ 0.003  & 20124 $\pm$ 4244, 6.7 $\pm$ 1.4  & 6693 $\pm$ 494, 2.2 $\pm$ 0.2 & 23928 $\pm$ 8840, 8 $\pm$ 3 \\ \rule{0pt}{3ex}
 $v_a$ [km/s] & 413.3 & 246.8 & 884.2 & 183.6 \\ \rule{0pt}{3ex}
 $e$                     & 0.890 $\pm$ 0.005   &  0.976 $\pm$ 0.01  & 0.768 $\pm$ 0.030  & 0.985 $\pm$ 0.011\\ \rule{0pt}{3ex}
 $t_{\rm period}$ [yr] & 15.9 $\pm$ 0.4 & 9.9 $\pm$ 0.3 & 7.6 $\pm$ 0.3 & 12.0 $\pm$ 0.3 \\
 \hline 
\end{tabular}
\caption{Stellar information for S2, S62, S4711 and S4714 \cite{2017ApJ...847..120H, Pei_ker_2020, Boehle_2016}. We list the stellar age, mass $M$, radius $R$, effective temperature $T_{\rm eff}$, log luminosity with solar luminosity ratio, pericenter distance $r_p$, apocenter distance $r_a$, pericenter velocity $v_p$ (in both km/s and percentage of the speed of light), the apocenter velocity $v_a$ (in km/s, calculated from elliptic orbit), the eccentricity $e$, and the orbital period $t_{\rm period}$. Some parameters are derived from others, e.g. $R \approx M^{0.8}$ for main sequence stars.}
\label{tab: star parameters}
\end{table*}

\subsection{Observations of Stars Near Sgr A*} 
Observations over the last 20 years have uncovered a young star cluster that lies extremely close to Sgr A*~\cite{Eckart:1997em, Ghez:2003qj, 2012A&A...545A..70S}. This ``S-Cluster" has at least 20 known members. The quintessential system, S2, has an observed mass of 13.6~$M_\odot$, and has a pericenter passage of Sgr A* of only 120~AU. This close proximity allows us to place strong constraints on the characteristics of the supermassive black hole~\cite{Ghez:2008ms, Gillessen:2008qv}. Spectroscopic observations indicate that the majority of observed S-Cluster stars are B-type main sequence stars, with putative ages of $\sim$1~Myr~\cite{Eisenhauer:2005cv}. Their close proximity to Sgr A* reveals a ``paradox of youth". Their low ages suggest that the stars were formed \emph{in situ} very close to Sgr A*~\cite{2017ApJ...847..120H}. However, standard models indicate that star-formation cannot occur so close to a central black hole, indicating that the stars should have migrated from larger radii~\cite{Ghez:2003qj}. More recent observations have discovered several less-luminous and less-massive stars that pass even closer to Sgr A*.

Table~\ref{tab: star parameters} lists the properties of the stars we focus on in our study: S2, S62, S4711, and S4714~\cite{2017ApJ...847..120H, Pei_ker_2020, Boehle_2016}. These stars have the closest average distance to Sgr A*, and span several mass ranges: S4711 and S4714 are both $\sim$2~$M_\odot$, S62 is $\sim$6.1~$M_\odot$, while S2 is the best-measured object with a mass of $\sim$13.6~$M_\odot$. Note that the detection and exact characteristics of the three stars S62, S4711, S4714 are currently debated, see e.g, Ref.~\cite{Pei_ker_2020, 2021A&A...645A.127G, 2021ApJ...918...25P, 2022A&A...657A..82G, 2022ApJ...933...49P}. In any case, our analysis of these stars is representative for S-cluster stars. The discovery of low-mass stars is especially beneficial, as previous studies have found that low-mass stars are more strongly affected by dark matter annihilation in their cores~\cite{1989ApJ...338...24S,Fairbairn:2007bn, Scott:2008ns}. Because each star has a high eccentricity (spanning 0.768 --- 0.985), we calculate the average dark matter interaction with each star by integrating over their 7 -- 16 year orbits. Note that the parameters in Tab.~\ref{tab: star parameters} are computed based on models that do not include dark matter annihilation, the addition of a new dark matter annihilation term may modify some model parameters, but typically only negligibly in the case that the star still appears as a main sequence star (an important exception is the age, which we will show can be much older with dark matter). In the case that dark matter disrupts the stellar evolution, these stars will not have the same properties as we will show shortly. In general we find that once dark matter annihilation begins to disrupt the star, the magnitude of change in the stellar parameters is so extreme that small variations in the specific attributes of the original star are not important for our constraints.

\subsection{Stellar Evolution Models}

We investigate the effect of additional energy injection due to dark matter annihilation on the evolution of our stars listed above, using the stellar modelling and evolution code \texttt{MESA}~\cite{2011ApJS..192....3P, 2013ApJS..208....4P, 2015ApJS..220...15P, 2018ApJS..234...34P, 2019ApJS..243...10P, 2023ApJS..265...15J}, version \texttt{r22.11.1} with SDK version \texttt{22.10.1}. We use the default \texttt{work} model and simulate stars with masses according to Tab.~\ref{tab: star parameters} for their main sequence phase, and \texttt{MESA} evolves the radius and luminosity accordingly. In our default models we assume an initial stellar metallicity value of $Z=0.02$. We have checked the impact of evolution also with extreme cases of $Z=0$ and $Z=0.1$, finding that our results on the maximum dark matter energy injection vary only by a factor of a few across the entire range.

To take into account the extra power created by the dark matter annihilation inside the star, we use the parameter \verb|inject_uniform_extra_heat| as well \verb|min_q_for_uniform_extra_heat| and \verb|max_q_for_uniform_extra_heat| to set a mass region inside the star where the extra power is injected, i.e. where the dark matter accumulates and annihilates. We consider injection in the stellar core, which we take to correspond to the inner 10\% of the star mass. We have also tested the effect of injection regions ranging through $1 - 100\%$ of the star mass, and find that such variations only have minor effects on the results. Specifically, for injection in the inner 1\% of the stellar mass, the change compared to injecting in the inner 10\% is less than a factor of 1, while at 100\%, the annihilation power a star can survive increases by a factor of about 2 -- 3. Especially as the bulk of dark matter scenarios lead to dark matter dominantly accumulating in the core, we expect the uncertainty from the dark matter annihilation position within the star to be small.

\section{Results and Discussion}
\label{sec:results}
Figure~\ref{fig:schematic} summarizes our findings. Dark matter accumulation and subsequent annihilation can provide a significant new energy source to the star in regions with high dark matter density. This can affect the star in three main ways:
\begin{enumerate}
    \item Pre-main sequence stars are particularly susceptible to extra sources of energy injection, and dark matter annihilation during this period can freeze the star in the Hayashi track (as shown for low-mass WIMP burners in Ref.~\cite{Scott:2008ns}, and previously estimated analytically in Ref.~\cite{1989ApJ...338...24S}), preventing them from ever forming and entering the main sequence. We define this freezing point within the \texttt{MESA} code, as the point when a star does not reach the main sequence even after 10 billion years (about the age of the Galaxy). 
    
    We note that we base our conclusion on the fact that \texttt{MESA} fails to evolve a star even after billions of years for the standard conditions, but different stellar collapse conditions might give varying results. We further note that we base our limits on prevented star formation on the scenario that a pre-main sequence star cannot move onto the main sequence, as \texttt{MESA} starts out the simulation with pre-main sequence stars. However, at similar dark matter densities, already earlier stellar evolution stages might be disrupted, thus even preventing the pre-main sequence star from forming.

    \item The stellar evolution is slowed down, as a fraction of the nuclear fusion needed to sustain the hydrostatic equilibrium of the star is replaced by the annihilating dark matter, resulting in the star using up its hydrogen more conservatively.
    
    \item Similarly, stars that are created in a region where the dark matter density is low, but migrate during their main sequence phase to a higher density region, can extend their lifetime, or be disrupted if the dark matter density is so high that the gravity of the star cannot counteract the dark matter annihilation power.
    
    Within the context of our models, a stellar disruption occurs when the star begins evolving backwards along the HR diagram and re-enters the pre main-sequence solution. This occurs due to the fact that the dark matter accumulation rate is constant, and once dark matter annihilation exceeds the stellar equilibrium energy and produces stellar expansion, there is no new equilibrium that can be established until the stellar gas is driven away from the star. We stress that this expansion will happen on timescales far longer than the timescales of standard time-domain astronomy searches, and the star will not ``explode" in a single timestep. We further note that, while the \texttt{MESA} solutions near disruption may be uncertain, this has little effect on our results, because disrupted stars will not look anything like observed S-cluster stars.
\end{enumerate}

\begin{figure}[tbp]
\centering
\includegraphics[width=0.48\textwidth]{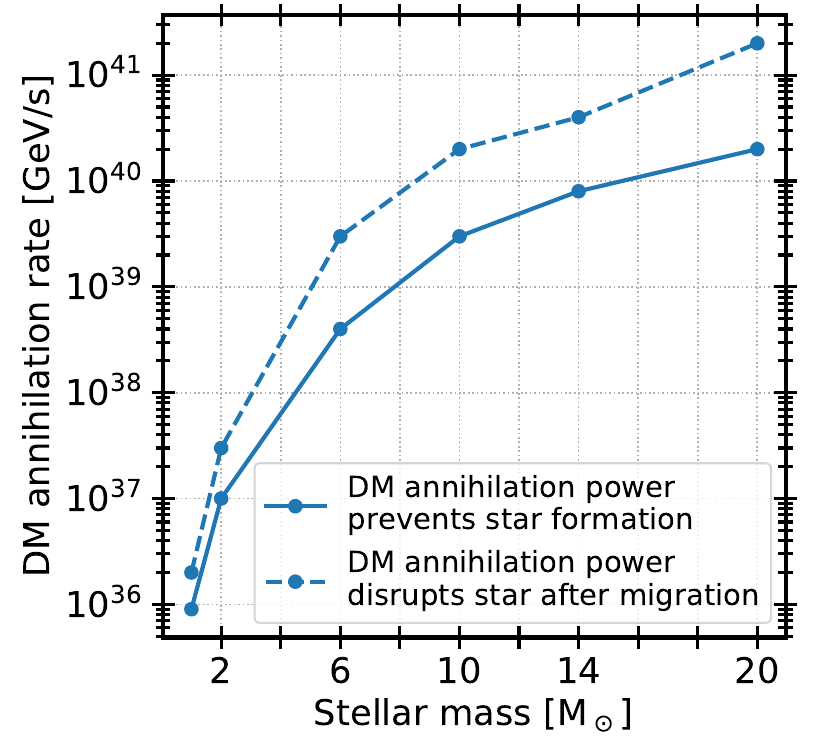}
\caption{The critical dark matter annihilation rate for two dramatic outcomes as labelled, as a function of stellar mass. Solid lines indicate where the dark matter annihilation power prevents star formation and dashed lines where the dark matter annihilation power disrupts the star after migration. Less massive stars are affected already at lower dark matter annihilation rates than more massive stars.}
\label{fig: DM annihilation rates vs stellar mass}
\end{figure}

Figure~\ref{fig: DM annihilation rates vs stellar mass} shows the critical dark matter annihilation rate for each of our scenarios as a function of the zero-age main sequence stellar mass. Solid lines represent the minimum dark matter annihilation rate at which the star is prevented from forming, and the dashed lines represent the minimum dark matter annihilation rate at which a star that successfully formed elsewhere and migrated inwards would be disrupted. Depending on the stellar mass, an annihilation rate of $\sim 10^{37}$~GeV/s is sufficient to prevent and disrupt a 1~$M_\odot$ star, while a more massive star of 20~$M_\odot$ survives dark matter annihilation power up to $\sim 10^{41}$~GeV/s.

\begin{figure*}[tbp]
\centering
\begin{minipage}[t]{0.48\textwidth}
\includegraphics[width=0.8\textwidth]{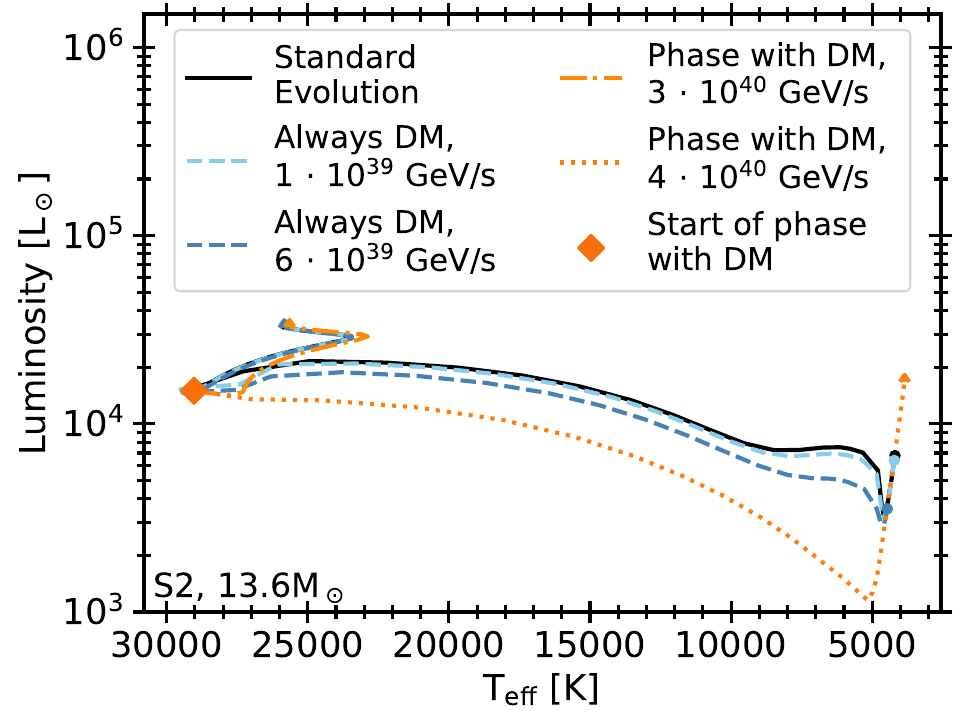}
\end{minipage}
\hfill
\begin{minipage}[t]{0.48\textwidth}
\includegraphics[width=0.8\textwidth]{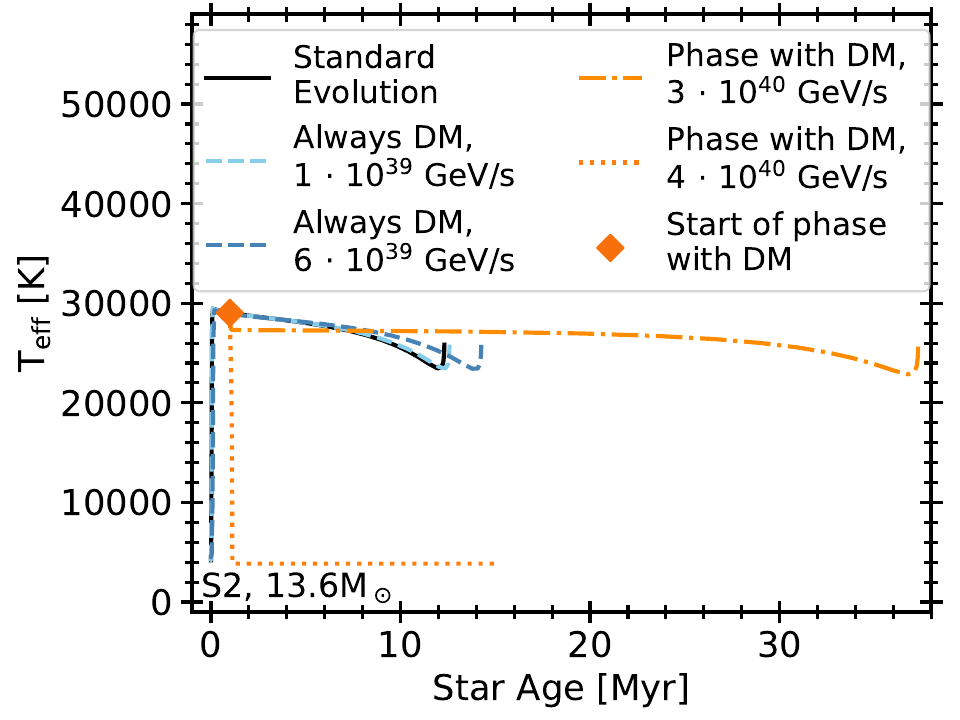}
\end{minipage}
\vfill
\begin{minipage}[t]{0.48\textwidth}
\includegraphics[width=0.8\textwidth]{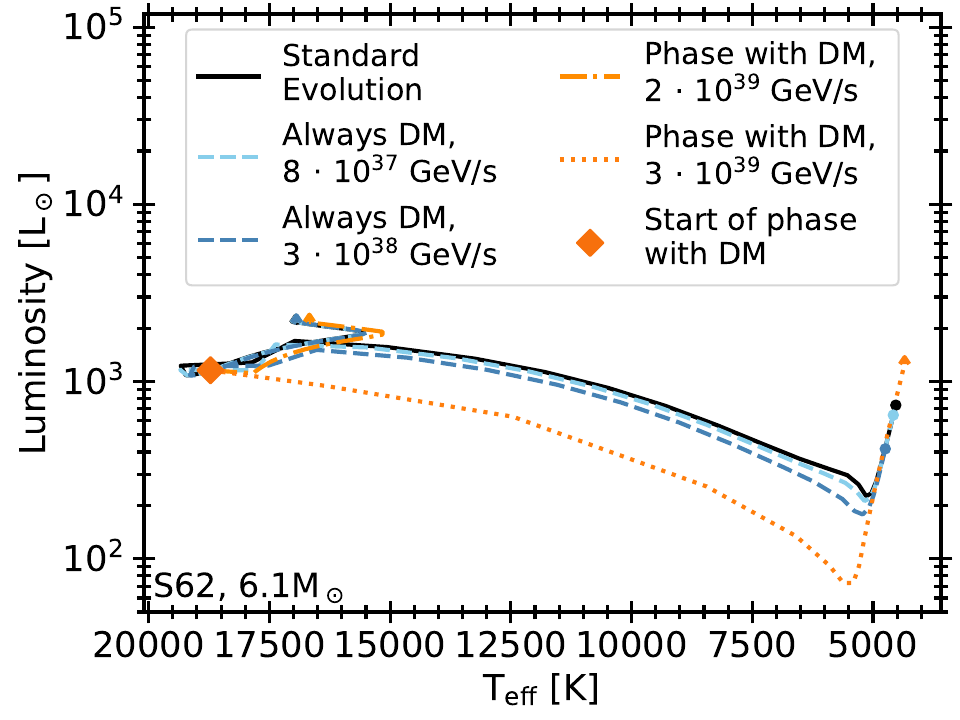}
\end{minipage}
\hfill
\begin{minipage}[t]{0.48\textwidth}
\includegraphics[width=0.8\textwidth]{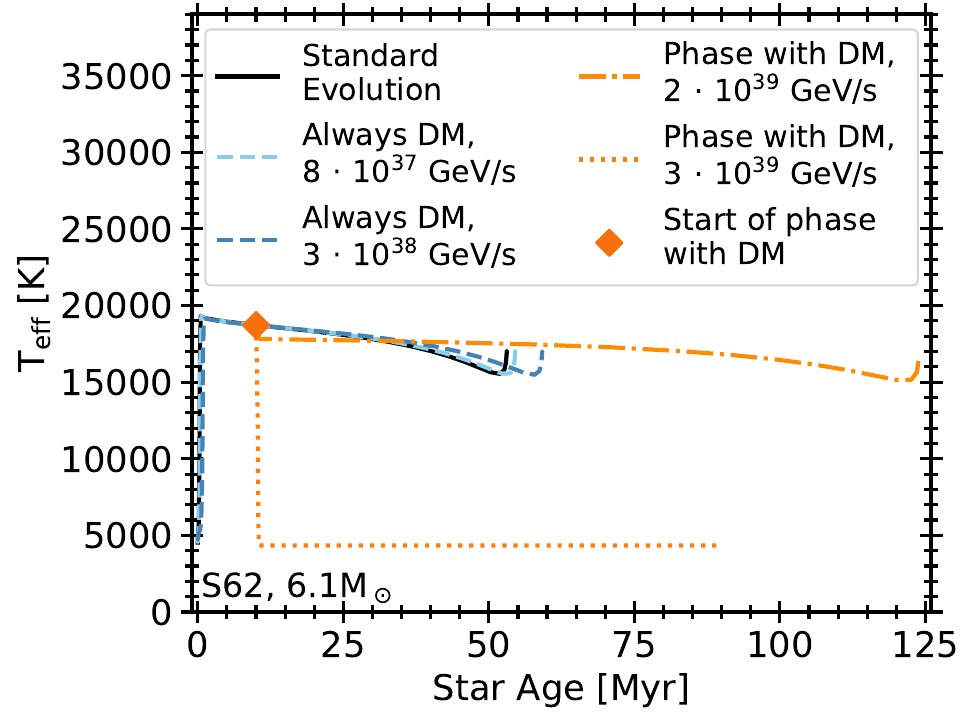}
\end{minipage}
\vfill
\begin{minipage}[t]{0.48\textwidth}
\includegraphics[width=0.8\textwidth]{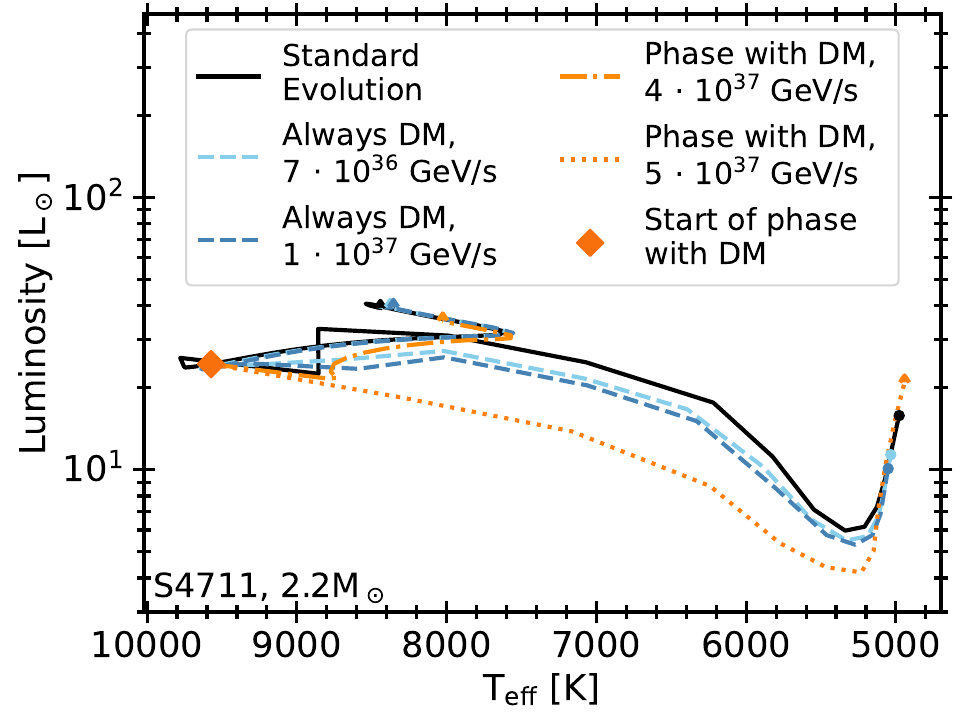}
\end{minipage}
\hfill
\begin{minipage}[t]{0.48\textwidth}
\includegraphics[width=0.8\textwidth]{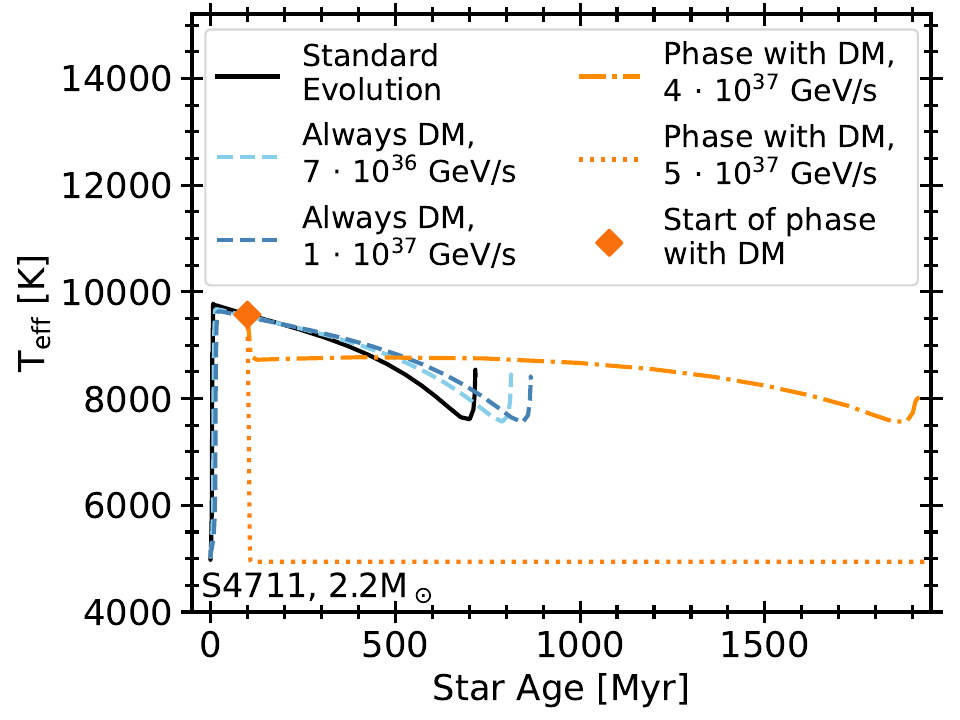}
\end{minipage}
\vfill
\begin{minipage}[t]{0.48\textwidth}
\includegraphics[width=0.8\textwidth]{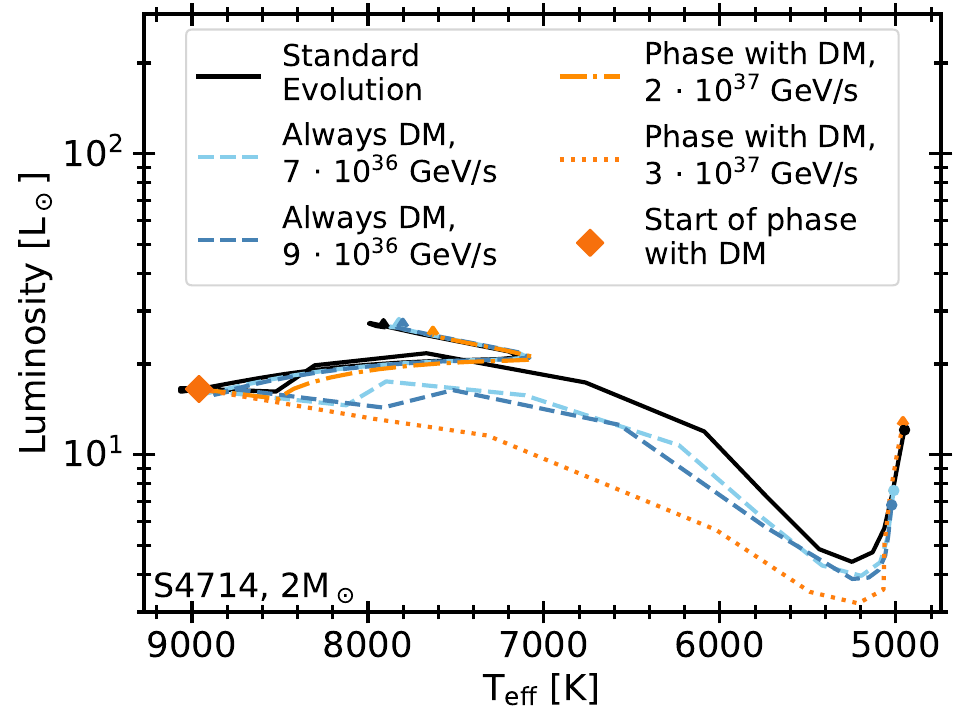}
\end{minipage}
\hfill
\begin{minipage}[t]{0.48\textwidth}
\includegraphics[width=0.8\textwidth]{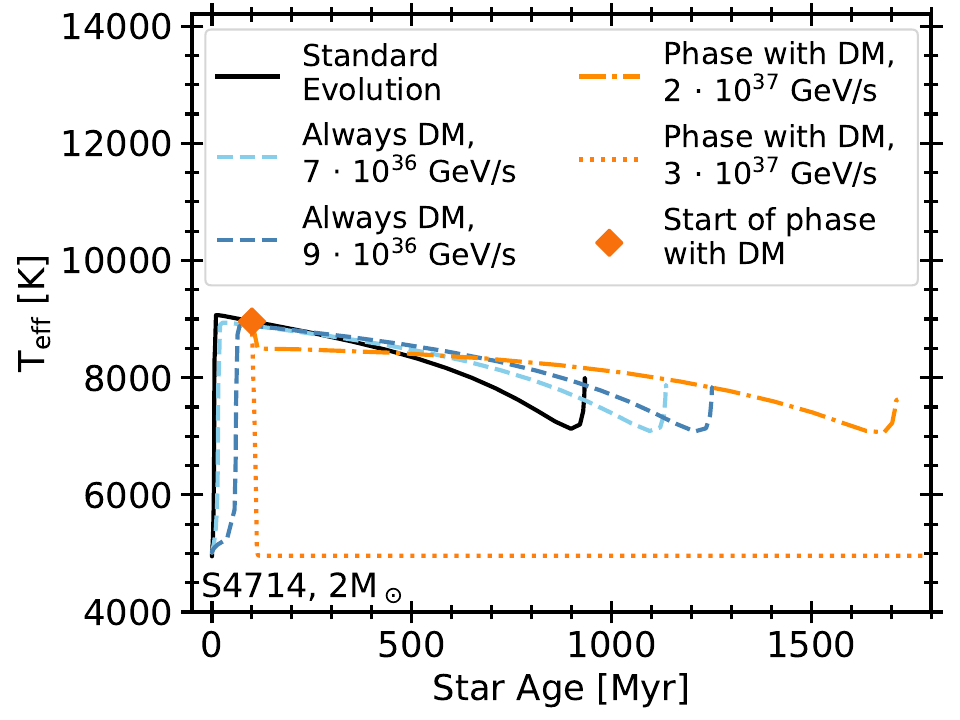}
\end{minipage}
\caption{The effect of dark matter on stellar evolution, for the stars S2, S62, S4711 and S4714 (top to bottom rows). For a given row, the left panel shows the HR diagram and the right panel shows the effective temperature against age for the labelled star. Black lines show the standard stellar evolution without dark matter, the other lines include dark matter annihilation with conditions as indicated in the legend. Blue lines show the case where dark matter is present from stellar birth for a smaller (light blue) and larger (dark blue) amount of dark matter. Orange lines represent stars that are affected by dark matter after they formed: First, the evolution follows the standard stellar evolution (black line), but at the orange diamond marker, dark matter is injected, and the evolution diverges from the standard scenario, for a high dark matter density (orange dot-dashed lines) and the disruption of the star when it migrates to a high-density region (orange dotted lines).}
\label{fig: HR diagrams}
\end{figure*}

\begin{figure*}[tbp]
\centering
\begin{minipage}[t]{0.42\textwidth}
\centering
\includegraphics[height=5.7cm, trim={0 0 4.8cm 0},clip]{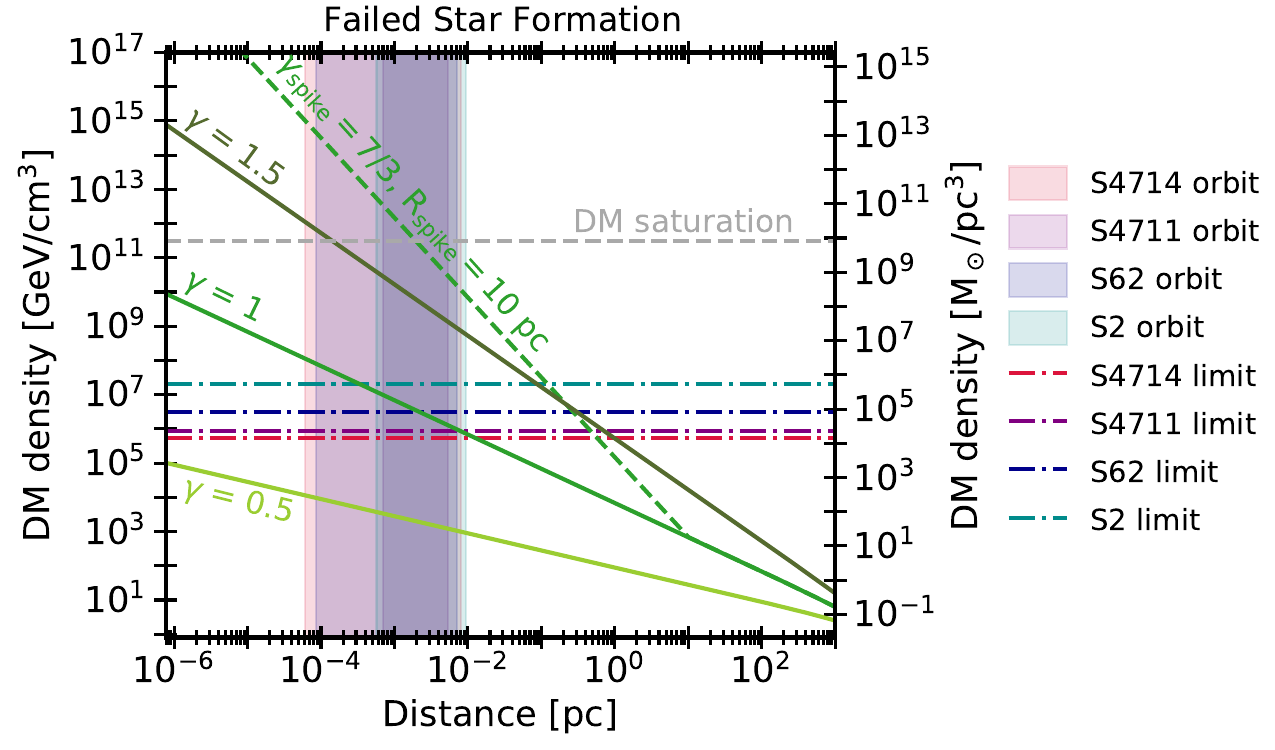}
\end{minipage}%
\begin{minipage}[t]{0.16\textwidth}
\centering
\includegraphics[height=5.5cm, trim={17cm 0 0cm 0},clip]{Plot_DM_profile_with_spike_center_failed_formation.pdf}
\end{minipage}%
\begin{minipage}[t]{0.42\textwidth}
\centering
\includegraphics[height=5.7cm, trim={0 0 4.8cm 0},clip]{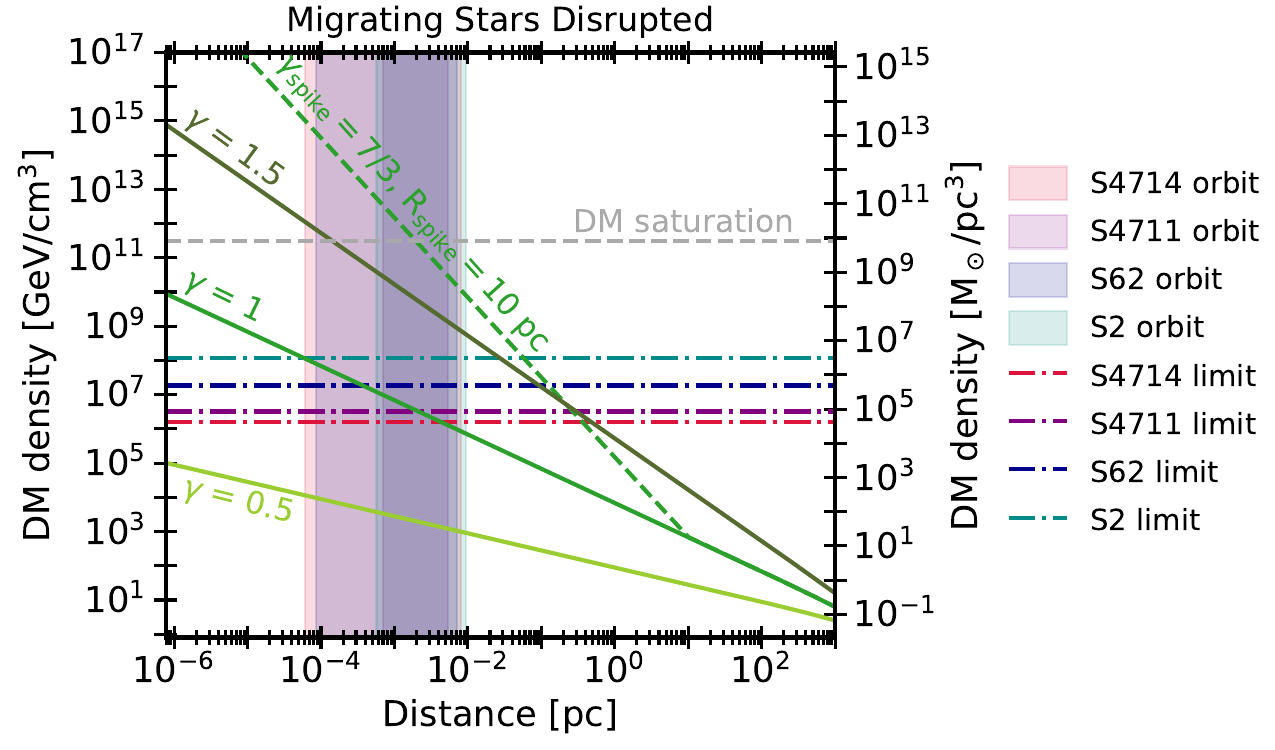}
\end{minipage}
\caption{The dark matter density as a function of Galactic Center distance for dark matter profiles with $\gamma = 0.5$ (light green), $\gamma = 1$ (medium green), $\gamma = 1.5$ (dark green), shown for models both with (dashed) and without (solid) a spike~\cite{Lacroix:2018zmg}. For the stars S2, S62, S4711 and S4714 we show the dark matter density above which dark matter annihilation prevents the star from forming (left panel) and above which a star that formed elsewhere and migrated to this region is disrupted (right panel). For reference, the distances covered by each stellar orbit are indicated in shaded bands. The dark matter saturation limit (grey, dashed) indicates the highest possible dark matter density before dark matter annihilation saturates the density (i.e. the spike would be flattened), here given as an example for a dark matter mass of 1~TeV and annihilation cross section of $10^{-26}$~cm$^3$/s.}
\label{fig: dark matter profiles}
\end{figure*}

Figure~\ref{fig: HR diagrams} HR diagrams (left), as well as the effective temperature as a function of the stellar age (right) for the stars S2, S62, S4711 and S4714 (from top to bottom respectively) for up to their main sequence phase. HR diagrams are useful to understand stellar evolution, and show the luminosity of the star against the stellar temperature (on a reversed $x$-axis). Newly forming stars start out at low temperatures and low luminosities (to the bottom right of the plot), but as hydrogen fusion is ignited in the stellar core, temperature and luminosity increase quickly, until the star reaches the main sequence, where temperature and luminosity barely change as the star undergoes hydrogen fusion in the core. When the hydrogen in the core is exhausted, hydrogen shell burning begins to maintain hydrostatic equilibrium, and the temperature decreases while the luminosity increases. When hydrogen shell burning becomes inefficient, at the next turnover point, the helium flash, helium burning starts in the core, and the star becomes a red giant, expanding hundred-fold in size. More evolutionary stages follow, but here, we focus on stars during their main sequence phase.

\begin{figure*}[tbp]
\centering
\begin{minipage}[t]{0.48\textwidth}
\centering
\includegraphics[width=1\textwidth, trim={0 0 10cm 0},clip]{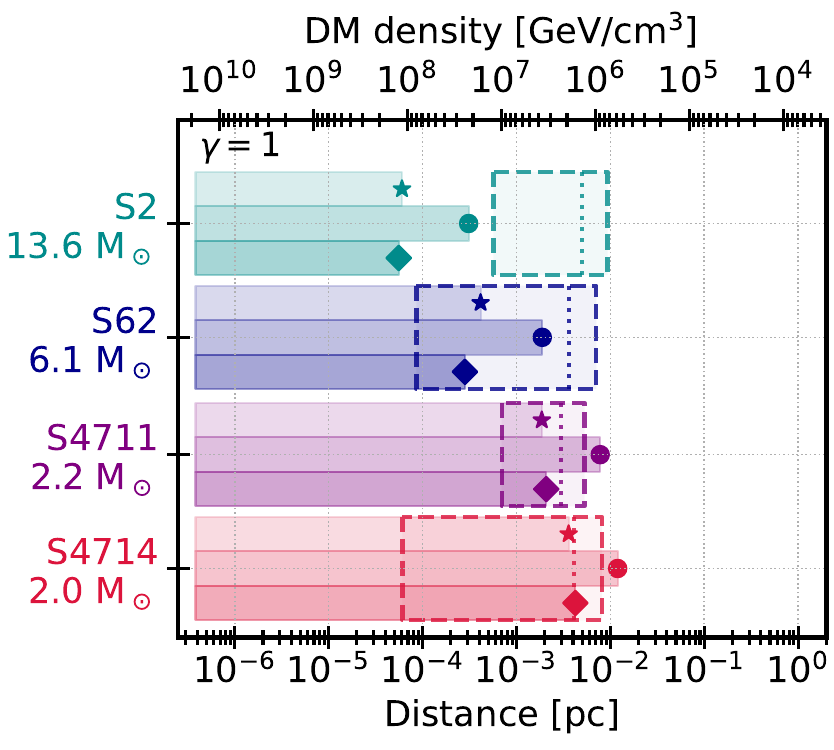}
\end{minipage}
\hfill
\begin{minipage}[t]{0.48\textwidth}
\centering
\includegraphics[width=1\textwidth, trim={0 0 10cm 0},clip]{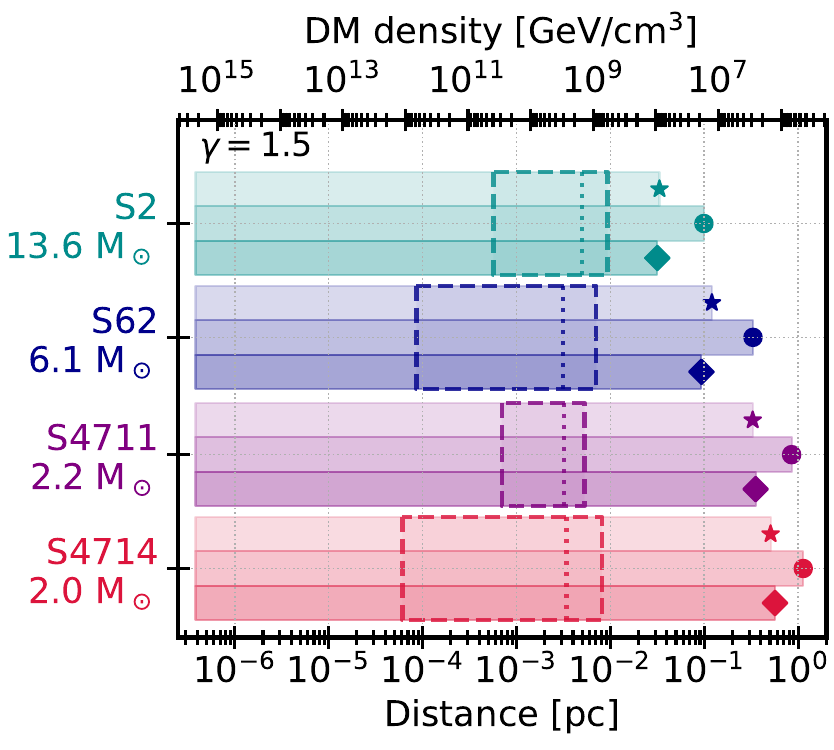}
\end{minipage}
\vfill
\begin{minipage}[t]{0.48\textwidth}
\centering
\includegraphics[width=1\textwidth, trim={0 0 10cm 0},clip]{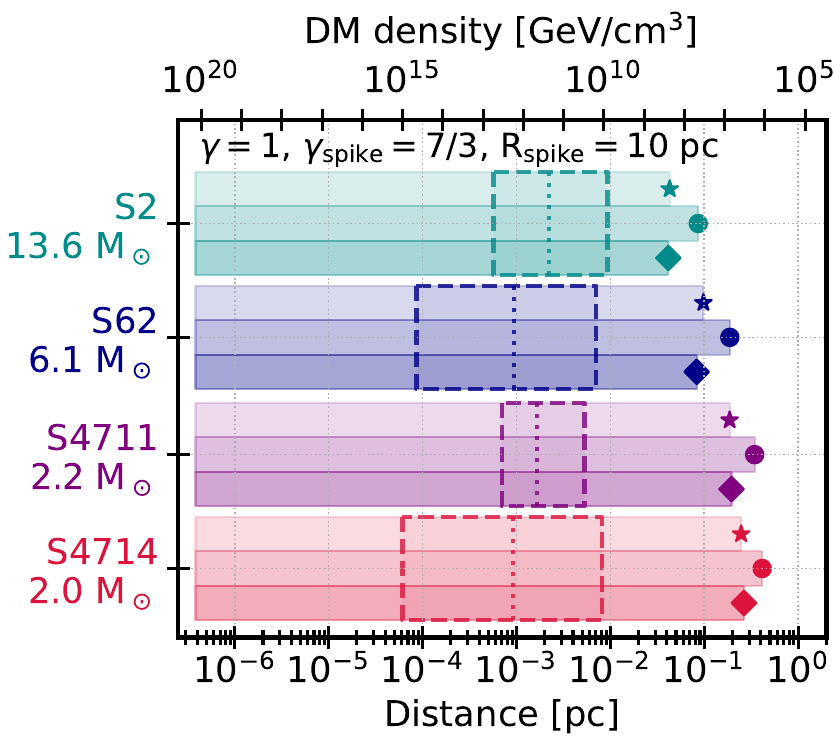}
\end{minipage}
\hfill
\begin{minipage}[t]{0.48\textwidth}
\centering
\includegraphics[width=0.65\textwidth, trim={14.1cm 0 0 0},clip]{Plot_Stars_vs_important_distances_gamma=1_with_spike.pdf}
\end{minipage}
\caption{The dark matter density (top axis) and distance to the Galactic Center (bottom axis) corresponding to certain events in the stellar evolution of our four stars. The shaded bars indicate the range of distances from the Galactic Center where three events occur: (1) the nuclear fusion power is exceeded by the dark matter annihilation power compared to a star without any dark matter (labelled with a star), (2) the dark matter annihilation power prevents the star from forming (circle), and (3) the dark matter annihilation power disrupts the star that formed further away and migrated to within this distance (diamond). Dashed boxes indicate the distance covered by the star's orbit with the dotted line representing the average dark matter density encountered along the orbit.  Constraints on the dark matter density profile can be obtained where the stellar orbit is within the shaded bars, i.e. where the dark matter annihilation dominates stellar processes and has a significant effect on the stellar evolution. We consider three dark matter profiles: NFW, $\gamma = 1$ (top left), $\gamma = 1.5$ (top right), $\gamma=1$ with spike $\gamma_{\rm spike} = 7/3$ and $R_\text{spike} = 10$~pc (bottom).}
\label{fig: stars vs distance}
\end{figure*}

In each plot of Fig.~\ref{fig: HR diagrams}, the standard stellar evolution (i.e. without dark matter) is shown in black. For the dark matter contribution, we consider two cases:

\begin{enumerate}
    \item Blue lines indicate the evolution in \emph{in situ} scenarios where the dark matter is already present during the pre-main sequence stages of stellar evolution. In this case, the lighter shade represents a smaller dark matter annihilation power, and the darker blue line includes a dark matter annihilation power that is just below the value that would prevent the star from forming.
    \item Orange lines indicate migration scenarios where a large dark matter annihilation power is only present after the star reaches the main sequence. By default, we choose to begin the phase with the dark matter annihilation power when the star is well within its main sequence phase, after 1~Myr for S2, 10~Myr for S62, and 100~Myr for S4711 and S4714. We have separately checked that a more gradual dark matter injection (which more closely reflects the migration process) gives comparable results. This is expected as disruption is only observed once the dark matter annihilation becomes comparable to or larger than nuclear fusion, and injecting dark matter below this threshold has little effect on the star. The phase before dark matter injection follows the black line, and the beginning of the dark matter phase is indicated by the orange diamond marker. The dark orange dash-dotted line shows a model where dark matter significantly increases the main sequence lifetime of the star. The orange dotted line shows the minimum dark matter injection energy that disrupts the star -- note that the star moves ``backwards" on the HR diagram towards lower luminosities and temperatures, as it is disrupted. For the exact extra power from the dark matter annihilation, see the legend in each plot.
\end{enumerate}

The results in Fig.~\ref{fig: HR diagrams} show how dark matter annihilation affects the evolution of the stars. While dark matter produces only small changes (both increases and decreases) in the luminosity and temperature of stars along the main sequence, the fact that dark matter provides a new energy source significantly slows down stellar evolution, making stars appear younger than they actually are. This may help relax the tension in the paradox of youth problem~\cite{Ghez:2003qj, Scott:2008ns, Eisenhauer:2005cv, Lu:2008iz, Hassani:2020zvz}, by giving stars extra time to migrate into the Galactic Center region. For example, in the case where S2 spends its entire lifetime in an environment that provides $6 \times 10^{39}$~GeV/s (Fig.~\ref{fig: HR diagrams} top panels, dark blue line), which is just below the annihilation rate that would prevent the star from forming, the main sequence lifetime is extended by about a factor of 15\%, from 12.4~Myr in the standard evolution scenario (black line) to 14.3~Myr in the dark matter case. In the maximum migration scenario, the lifetime is even more extended. For a dark matter annihilation power of $3\times 10^{40}$~GeV/s, just below the power that would disrupt the star, S2 does not exit the main sequence until 37.3~Myr, which is a factor of 3 slower than the scenario without dark matter annihilation. The results shown in Fig.~\ref{fig: HR diagrams} are based on the specific dark matter annihilation power that is needed to noticeably affect the stellar evolution, but it is not necessary to assume a specific dark matter model (i.e. dark matter mass, dark matter-nucleon scattering cross section and dark matter density). Rather, any model that reproduces the annihilation power is valid.

\begin{figure*}[tbp]
\centering
\begin{minipage}[t]{0.48\textwidth}
\centering
\includegraphics[width=\textwidth]{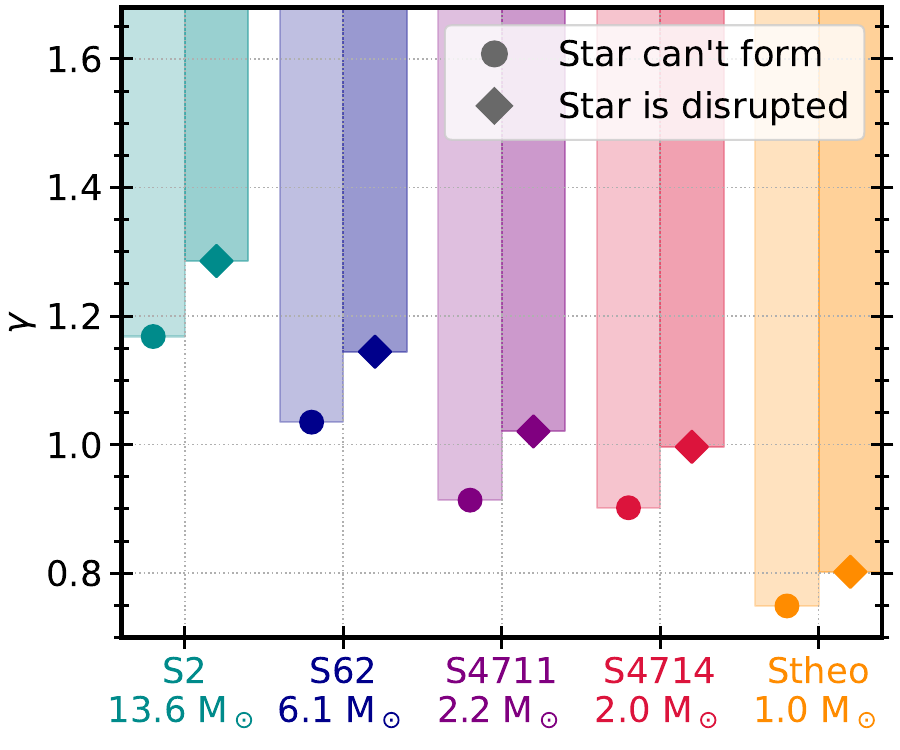}
\caption{Exclusion limits on the dark matter profile (see Eq.~(\ref{eq: DM profile})) for prevented star formation (circle) and star disruption after migration (diamond) based on the existence of the four stars S2 (teal), S62 (dark blue), S4711 (purple) and S4714 (red). We also show a hypothetical 1~$M_\odot$ star (orange) with an orbit twice as close to Sgr A* than S4714.}
\label{fig: DM profile constraints}
\end{minipage}
\hfill
\begin{minipage}[t]{0.48\textwidth}
\centering
\includegraphics[width=\textwidth]{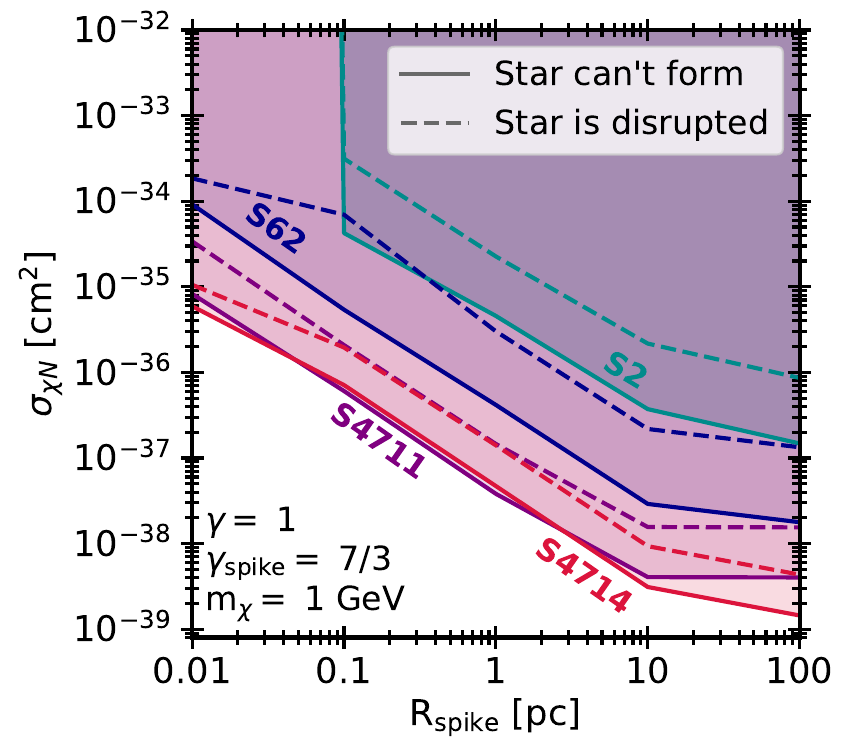}
\caption{The dark matter--nucleon cross section for different spike dark matter profiles (see Eq.~(\ref{eq: DM profile with spike})) for varying $R_\text{spike}$ and $\gamma=1$, $\gamma_\text{spike}=7/3$ for a fixed dark matter mass of $m_\chi = 1$~GeV, for prevented star formation (solid lines) and star disruption after migration (dashed lines) based on the existence of the four stars S2 (teal), S62 (dark blue), S4711 (purple) and S4714 (red).}
\label{fig: DM profile constraints spike}
\end{minipage}
\end{figure*}

Figure~\ref{fig: dark matter profiles} shows the dark matter density as a function of the distance from Sgr~A* for different dark matter profiles. Solid lines indicate the standard dark matter profile given in Eq.~(\ref{eq: DM profile}) for different profile indices $\gamma$, while dashed lines represent the spike model given in Eq.~({\ref{eq: DM profile with spike}}). The gray dashed line shows the saturation limit (see Eq.~(\ref{eq: DM density saturation})) for an example 1~TeV annihilating dark matter particle with an annihilation cross section of $10^{-26}$~cm$^{3}$/s. Above this limit, the dark matter density cannot increase because it is counteracted by the dark matter annihilation rate. The dashed-dotted lines indicate the dark matter density at which the dark matter annihilation power is so high that the stars are prevented from forming (left panel) and stars that successfully formed elsewhere are disrupted after moving towards the inner region with higher dark matter density (right panel) for S2 (teal), S62 (dark blue), S4711 (purple) and S4714 (red). The shaded bands cover the stellar orbits. For this figure, we assume the geometric dark matter capture rate (i.e., all dark matter particles that pass through the star are captured).

Figure~\ref{fig: stars vs distance} further demonstrates the stellar survival given a dark matter profile. The bottom axis shows the distance from the Galactic Center, and the top axis shows the corresponding dark matter density predicted by the chosen dark matter profile for that plot. For each star, we indicate where the dark matter annihilation power: (1) corresponds to the nuclear fusion power of a star that has no dark matter contribution (star marker), (2) prevents the star from forming (circle marker), and (3) disrupts the star (diamond marker). The dashed boxes cover the stellar orbit and the vertical dotted lines indicate the average dark matter density throughout the orbit. Typically, stellar formation is prevented when the dark matter annihilation rate reaches a fraction of the nuclear fusion power, and the star is disrupted when the dark matter annihilation power is comparable to the nuclear fusion power. Note that in a star, the dark matter power may, at times, be slightly stronger than the nuclear fusion power without disrupting the star.

In Fig.~\ref{fig: stars vs distance}, we consider various dark matter profiles. In the top left panel ($\gamma = 1$), the stellar orbit of S2 does not reach the distance where the dark matter density is high enough to disrupt the star, and S62 and S4711 are only somewhat affected. Only S4714's orbit covers a dark matter density high enough to reach the disruption limit. In the top right panel ($\gamma = 1.5$) and bottom left panel (spike model with $\gamma_\text{spike} = 7/3$ and $R_\text{spike} = 10$~pc), all stellar orbits, and the average dark matter density that they encounter, are well within the formation and disruption limit where they are dominated by the dark matter annihilation power. This means that at these dark matter densities and distances to Sgr A*, these stars would not be able to exist. Note that these constraints are dependent mainly on two stellar parameters: the stellar mass, and the stellar orbit. The stellar mass determines how much extra power from dark matter annihilation a star can survive, with more massive stars being able to withstand higher densities. The stellar orbit determines how much dark matter is captured for a specific profile. In this figure, we assume that effectively all dark matter passing through the star is captured, i.e. the maximum possible capture rate.

Figure~\ref{fig: DM profile constraints} summarizes constraints on the dark matter profiles for different values of $\gamma$, assuming no dark matter spike (see~Eq.~(\ref{eq: DM profile})), and assuming that all the incoming dark matter is captured. While the weakest constraints are obtained for S2, excluding profiles above ${ \gamma \gtrsim 1.3} $, the strongest constraints are given by S4714, the star that is lightest and reaches closest to Sgr A*, excluding all profiles above $\gamma \gtrsim 1.0$. Even stronger limits can be obtained assuming that the stars would form in the region of their current orbits, which excludes all profiles with $\gamma \gtrsim 0.9$. Additionally, we show the expected limit for a hypothetical star with a mass of 1~$M_\odot$ and an orbit twice as close to Sgr A* as S4714. This star would only survive dark matter annihilation rates about an order of magnitude lower than S4711 and S4714 and encounters a higher dark matter density around its orbit due to its closer proximity to the Galactic Center. The discovery of such a star would provide constraints on the dark matter profile reaching below $\gamma \gtrsim 0.8$. All spike models are excluded by at least S4714 for the parameters here considered (${ \gamma = 1 }$, ${ \gamma_\text{spike} = 7/3 }$ and $R_\text{spike} = 0.01 - 100$~pc).

Figure~\ref{fig: DM profile constraints spike} shows constraints on the spike dark matter profile with $\gamma = 1$, $\gamma_\text{spike} = 7/3$ and ${ R_\text{spike} = 0.01 - 100 }$~pc (see~Eq.~(\ref{eq: DM profile with spike})).
We present the dark-matter--nucleon scattering cross section against $R_\text{spike}$ for a dark matter particle with mass $m_\chi = 1$~GeV. Limits based on failed star formation are given in solid lines and limits based on stellar disruption after migration in dashed lines. We include a dark matter density saturation (see Eq.~(\ref{eq: DM density saturation})) with $m_\chi = 1$~GeV and $\langle\sigma v\rangle = 10^{-26}$~cm$^3$/s; if dark matter has instead dominantly $p-$wave annihilation, these bounds would be stronger due to the larger dark matter spike not being depleted.

\begin{figure*}[tbp]
\centering
\begin{minipage}[t]{0.495\textwidth}
\centering
\includegraphics[width=1\textwidth]{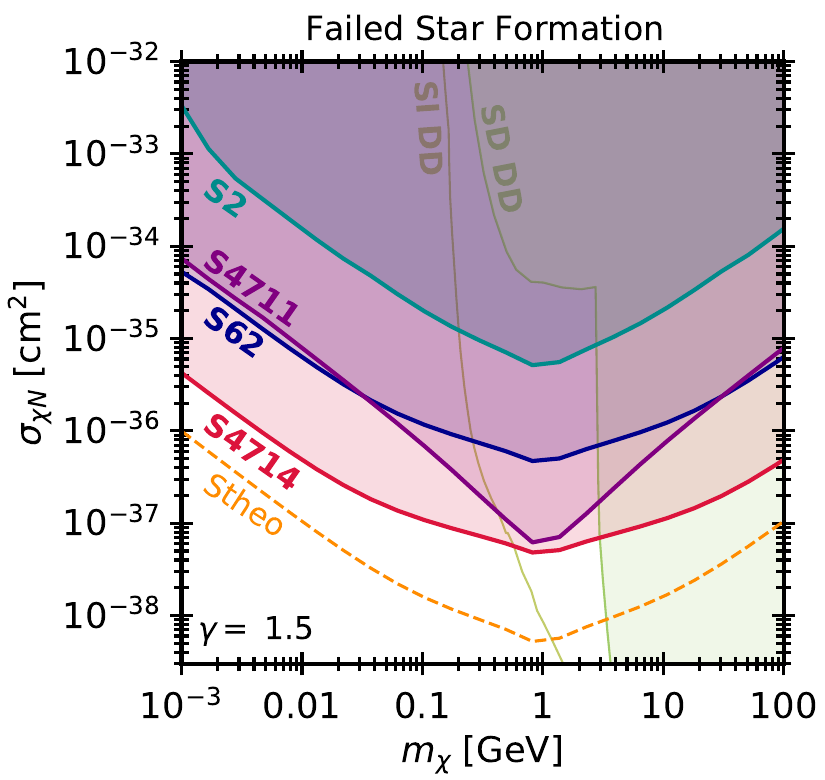}
\end{minipage}
\hfill
\begin{minipage}[t]{0.495\textwidth}
\centering
\includegraphics[width=1\textwidth]{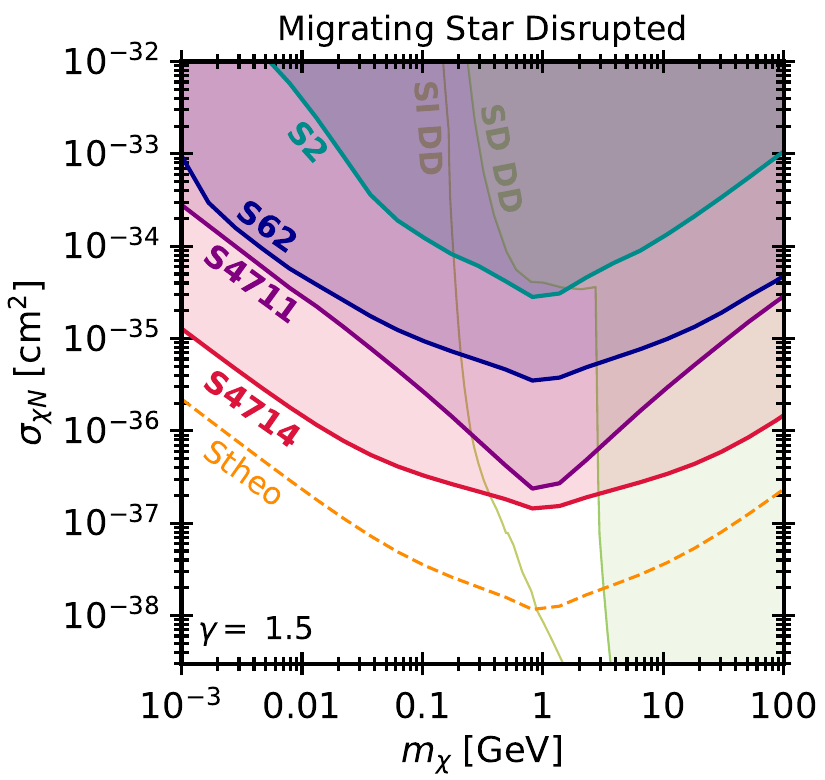}
\end{minipage}
\caption{Constraints on the dark matter-nucleon scattering cross section against the dark matter mass as derived from failed star formation (assuming \textit{in situ} formation) (left panel) and disruption of migrated stars (right panel) for our four S-stars: S2, S4711, S62, S4714. For comparison we show complementary constraints, which arise from direct detection experiments from either spin-independent ``SI DD" or spin-dependent ``SD DD" scattering; whether these are applicable will depend on the dark matter particle model. Additionally, we show projections for a hypothetical 1~$M_\odot$ star (orange) with an even closer orbit around Sgr A*, which would exceed our constraints from observed stars by another order of magnitude.}
\label{fig: cross section vs DM mass limits}
\end{figure*}

Figure~\ref{fig: cross section vs DM mass limits} displays our constraints on the dark matter-nucleon scattering cross section as a function of dark matter mass. We show constraints for each of the four stars based on failed star formation (left panel) and stellar disruption after migration (right panel), for a dark matter profile with $\gamma = 1.5$. We obtain the strongest cross section constraints for S4714, which is the lightest star with the most eccentric orbit, reaching closest to the Galactic Center. We constrain cross sections down to $5.9 \times 10^{-38}$~cm$^2$ based on preventing \emph{in situ} star formation and $1.7 \times 10^{-37}$~cm$^2$ based on the disruption of migrated stars. We also show the expected limits for the hypothetical 1~$M_\odot$ star (orange) with an orbit twice as close to Sgr A* than S4714. A discovery of such a star would improve our constraints by about an order of magnitude. Overall, we see that our most stringent limits arise from the fact that dark matter scattering and annihilation predicts failed stellar formation. However it is important to note that it is currently possible that all these stars migrated inwards from lower densities after forming elsewhere, and so our constraints from stellar disruption are more robust. If migration scenarios are ruled out in the future, the stronger \emph{in situ} constraints will be applicable.

In Fig.~\ref{fig: cross section vs DM mass limits}, we show complementary constraints from direct detection experiments on Earth. We display limits for the spin-independent dark matter-nucleon scattering cross section from CRESST-III~\cite{CRESST:2019jnq}, DarkSide~\cite{DarkSide:2018bpj}, XENON-nT~\cite{XENON:2023cxc}, and LZ~\cite{LZ:2022ufs}. We also show limits on the spin-dependent cross section (assuming pure interactions with protons) from CRESST-III~\cite{CRESST:2022dtl} and PICO-60~\cite{PICO:2019vsc}. As we have assumed the simplifying case of pure hydrogen in our stars, our stellar limits do not change between spin-independent and spin-dependent scattering, and only the complementary direct detection bounds will change depending on whether the dark matter model has spin-independent or spin-dependent scattering. We see that our constraints cover more new parameter space in the spin-dependent scenario, but that we also deliver new sensitivities for spin-independent models. There are also classes of dark matter models where these direct detection limits may not apply, but we show them regardless to facilitate comparison.

In Fig.~\ref{fig: cross section vs DM mass limits}, we do not include any dark matter evaporation, as it is highly model dependent~\cite{Acevedo:2023owd}. Evaporation truncates the lightest dark matter mass that is probed by our search, due to the fact that the thermal kicks imparted to the dark matter become too large given the gravitational potential energy of the star, allowing the dark matter to simply leave the system. While the purpose of our paper is not to investigate any detailed particle dark matter models, we provide for reference some evaporation cut offs in benchmark scenarios. In the case that the dark matter scattering is purely via contact interactions, we find that the evaporation mass in our parameter space is about a GeV, depending on the star and the cross section. In the case of attractive long-range interaction model classes, the evaporation mass can be sub-MeV~\cite{Acevedo:2023owd}. In a similar vein, including enhanced dark matter capture rates expected from attractive long-range particle models could increase our cross section sensitivity by orders of magnitude, depending on the specific model parameters. However to avoid detailing any specific model, and to be conservative, we only show cross section limits under the assumption of capture via purely contact interactions.

We note that the constraints that we have derived here operate under the assumption that the S-cluster stars are typical main sequence stars. This assumption is compatible with current observational constraints on S-star stellar parameters~\cite{Ghez:2003qj, Martins:2007rv,  
 Ghez:2008ms, Gillessen:2008qv, 2010RvMP...82.3121G, 
2017ApJ...847..120H, Pei_ker_2020}, but it is possible that the extreme environment near Sgr A* produces stars with extremely different physical characteristics than assumed in standard stellar evolution models (e.g., \texttt{MESA}). While this is not likely given current observations, if future scenarios strongly affect the structural parameters of these stars, the limits in this work would need to be re-evaluated.

\section{Summary and Outlook}
\label{sec:conc}

In this \textit{paper}, we have shown that stars in close proximity to Sgr A* are strongly affected by the high dark matter density in the innermost 10$^{-3}$~pc of the Galactic Center. Dark matter that is efficiently captured and annihilates in the stellar core can provide significant extra power to the star. Specifically, we have found that (1) stars can be prevented from forming (i.e. ever reaching the main sequence), (2) the stellar evolution is slowed down, making stars appear younger, and (3) stars that successfully form in a low-density region, and migrate during their main sequence phase to higher density regions in the Galactic Center can be disrupted.

We have based our study on the observation of four S-stars spanning a variety of stellar masses, closely orbiting Sgr A* on highly eccentric orbits. Within the context of standard WIMP dark matter models, the existence of these stars sets constraints on the dark matter density profile, as a sufficiently high density would disrupt the stars. Under the assumption that essentially all incoming dark matter is captured, we exclude profiles above $\gamma \sim 1.0$, i.e. we exclude standard NFW profiles or steeper. Note that when assuming dark matter masses and dark-matter-nucleon scattering cross sections for which the capture is not saturated, these constraints can be weakened accordingly. Furthermore, for a range of dark matter capture fractions, we place limits on the dark-matter-nucleon scattering cross section, that exceed results from direct detection experiments by several orders of magnitude for dark matter masses below $\sim 3$~GeV.

Our results show that lighter stars are disrupted at much lower dark matter densities than more massive stars. Notably, the future discovery of a star lighter than $\sim 2$~$M_\odot$ and/or on an orbit even closer to Sgr A* can place stronger constraints on the dark matter profile, as demonstrated by our hypothetical 1~$M_\odot$ star. While the dark matter accumulation rate also depends on the stellar size, velocity and orbit, more massive stars can generally survive a higher dark matter density and could therefore exist much closer to the Galactic Center than lighter stars, resulting in a gradient of star masses corresponding to the dark matter density profile.

We note that we assume that these S-cluster stars are main sequence stars and that the stars in our scenarios (1) and (3) (i.e. stars prevented from evolving into main sequence stars, and stars that migrate towards the Galactic Center as main sequence stars but are disrupted once they experience high enough dark matter densities) would be observationally distinguishable from a main sequence star.

Notably, our result may provide a solution to the paradox of youth problem. In models with intermediate dark matter densities, stars would be prevented from forming \emph{in situ} near Sgr A*. However, stars that migrate into the Galactic Center region from further distances may be much older than expected, weakening the tension in the paradox of youth problem~\cite{Ghez:2003qj, Scott:2008ns, Eisenhauer:2005cv, Lu:2008iz, Hassani:2020zvz}.

Finally, we point out that the unexpected paucity of old stars in the Galactic Center (known as the ``conundrum of old age")~\cite{Buchholz_2009, Merritt_2010}, as well as the missing pulsar problem~\cite{Dexter:2013xga}, may be explained if main sequence stars are eventually disrupted during later stages of their stellar evolution. We will explore this further in an upcoming publication.

\section*{Acknowledgments}
We thank Jeff Andrews, Djuna Croon, Joakim Edsj\"o, Andrea Ghez, Jeremy Sakstein, and Juri Smirnov for helpful discussions. I.J. and T.L. acknowledge support by the Swedish Research Council under contract 2022-04283. T.L. also acknowledges support from the European Research Council under grant 742104 and the Swedish National Space Agency under contract 117/19. R.K.L. is supported by the U.S. Department of Energy under Contract DE-AC02-76SF00515. This project used computing resources from the Swedish National Infrastructure for Computing (SNIC) and the National Academic Infrastructure for Supercomputing in Sweden (NAISS) under project No. 2022/3-27, partially funded by the Swedish Research Council through grant no. 2018-05973.

\bibliography{main}
\end{document}